\documentstyle[11pt,newpasp,twoside,epsfig]{article}
\index{summary}
\index{instructions}
\index{template}

\def\edcomment#1{\iffalse\marginpar{\raggedright\sl#1\/}\else\relax\fi}
\reversemarginpar

\begin{document}
\title{Observations of the Non-Thermal X-ray Emission from the Galactic 
Supernova Remnant G347.3-0.5}
\author{Thomas G. Pannuti}
\affil{MIT Center for Space Research, 77 Massachusetts Avenue, NE80-6015,
Cambridge, MA 02139}
\author{Glenn E. Allen}
\affil{MIT Center for Space Research, 77 Massachusetts Avenue, NE80-6029,
Cambridge, MA 02139}

\begin{abstract}
G347.3-0.5 (RX J1713.7-3946) is a member of the new class
of shell-type Galactic supernova remnants (SNRs) that feature
non-thermal components to their X-ray emission. We have analyzed 
the X-ray spectrum of this SNR over a broad energy range (0.5 to
30 keV) using archived data from observations made with 
two satellites, the R\"{o}ntgensatellit ($\it{ROSAT}$) and the 
Advanced Satellite for Cosmology and Astrophysics ($\it{ASCA}$), 
along with data from our own observations made with the Rossi X-ray
Timing Explorer ($\it{RXTE}$). Using a combination of the models
EQUIL and SRCUT to fit thermal and non-thermal emission, respectively,
from this SNR, we find evidence for a modest thermal component to
G347.3-0.5's diffuse emission with a corresponding energy of $\it{kT}$
$\approx$ 1.4 keV. We also obtain an estimate of 70 TeV for the
maximum energy of the cosmic-ray electrons that have been accelerated
by this SNR. 
\end{abstract}

\section{Introduction}
The X-ray luminous Galactic supernova remnant (SNR) G347.3-0.5 (RX 
J1713.7-3946) was discovered during the R\"{o}ntgensatellit ($\it{ROSAT}$)
All-Sky Survey (Pfeffermann \& Aschenbach 1996). Subsequent 
studies of this SNR's X-ray properties (Koyama et al. 1997; Slane
et al. 1999) revealed that most of its X-ray emission is non-thermal
and very likely produced by the synchrotron process. G347.3-0.5
therefore becomes a member of a new class of young shell-type 
SNRs that feature non-thermal components to their X-ray emission.
Other members of this class include Cas A (Allen et al. 1997),
SN 1006 (Koyama et al. 1995, Allen et al. 2001) and G266.2-1.2
(Slane et al. 2001). TeV gamma rays have been detected from the
X-ray luminous northwestern rim of this SNR (Muraishi et al. 2000),
making G347.3-0.5 only the third SNR (besides SN 1006 and Cas A)
where such high-energy emission is detected. The presence of this
emission suggests that acceleration of cosmic-ray electrons is taking
place along this rim, and that additional study of this SNR's X-ray
emission may lead to new insights on how SNRs act as cosmic-ray
particle accelerators.
\par
In order to analyze the X-ray properties of G347.3-0.5 in more detail,
as well as to study how cosmic-ray particles are accelerated by this 
SNR, we observed this source using the Rossi X-ray Timing
Explorer ($\it{RXTE}$). We supplemented the data from these
observations with publicly available data from observations that 
were made of G347.3-0.5 by two other X-ray satellites, $\it{ROSAT}$ and the
Advanced Satellite for Cosmology and Astrophysics ($\it{ASCA}$). 
Parameters for the X-ray observations used in this analysis are listed
in Table 1. By combining the data from all three satellites, we have 
sampled the X-ray emission from this SNR over the energy range of 0.5
through 30 keV.

\begin{table}
\caption{Summary of X-ray Observations of G347.3-0.5}
\begin{tabular}{lccccc}
\tableline
& & Observed & RA & Dec & Exposure \\
& & Portion of & (J2000.0) & (J2000.0) & Time \\
Satellite & Instrument & G347.3-0.5 & (h m s) & 
(\deg~\arcmin~\arcsec) & (Seconds)\\
\tableline
{\it ROSAT} & {\it PSPC} & All & 17 13 33.60 & $-$39 48
36.0 & 2758\\
{\it ASCA} & {\it GIS} & NW Rim & 17 12 17.76 & $-$39 35
30.8 & 20337\\
& & SW Rim & 17 12 53.76 & $-$39 54 23.4 & 18562\\
& & NE Region & 17 14 28.32 & $-$39 35 26.2 & 16220\\
& & SE Region & 17 15 41.52 & $-$40 02 25.8 & 40153\\
{\it RXTE} & {\it PCA} & All & 17 14 11.04 & $-$39 50 31.2 & 
45000\\
\tableline
\tableline
\end{tabular}
\end{table}

\section{Analysis}

Previous analyses of the X-ray emission from G347.3-0.5 have 
considered only the $\it{ROSAT}$ observations (Pfeffermann
\& Aschenbach 1996) or a combination of the $\it{ROSAT}$ and
$\it{ASCA}$ observations (Slane et al. 1999). In the latter
work, spectral fits were made to the X-ray luminous northwestern
and southwestern rims as well as the entire SNR, and simple
power-laws were found to adequately model both the emission from
the rims as well as from the SNR itself. 
\par
After extracting spectra for the SNR as observed by all three
satellites, spectral fitting was performed using the XSPEC
software package. We first tried fitting the whole X-ray 
spectrum with the power-law fits obtained by Slane et al. 1999.
We also used the SRCUT model (Reynolds 1996, Reynolds
\& Keohane 1999) to fit the data: this model fits a synchrotron spectrum 
from an exponentially cut-off power-law distribution of electrons
in a uniform magnetic field. The corresponding relativistic
electron energy spectrum can be expressed as
\begin{equation}
N_e (E) = K E^{-(2\alpha+1)}e^{\frac{-E}{E_{max}}}, 
\end{equation}
where $K$ is a normalization constant derived from the observed
flux density of the SNR at a frequency of 1 GHz, $\alpha$ is
the observed radio spectral index of the SNR and $E$$_{max}$ is
the cut-off electron energy. A crucial advantage of the SRCUT
model is that two of the parameters -- $K$ and $\alpha$ -- are
constrained by radio observations. In addition, the model returns
a value for the break frequency $\nu$$_{break}$ of the electron energy 
spectrum, and an estimate for the maximum energy of the accelerated
electrons can be made based on the value for this frequency.
\par
Lastly, we note that G347.3-0.5 lies within the Galactic Ridge,
which contributes a considerable amount of background emission to the 
$\it{RXTE}$ observations. We modeled this emission using a two-component
model (RAYMOND-SMITH and POWER LAW) with the parameters determined
by Valinia \& Marshall (1998) from their $\it{RXTE}$ observations
of this region (${\it kT}$ = 2.9 keV and $\Gamma$ = 1.8). To help 
improve the fit and search 
for thermal emission from the SNR, we included a thermal equilibrium
ionization model (EQUIL) in the fitting process. See Figure 1 for
the fit to the SNR's diffuse emission.
\begin{figure} 
\psfig{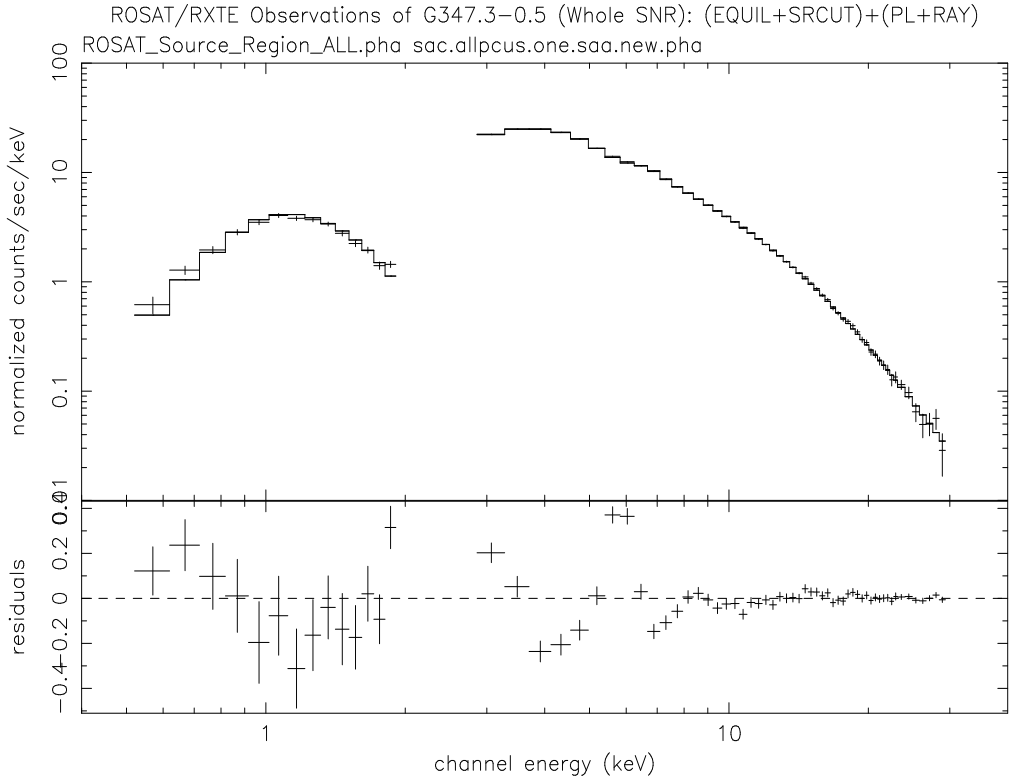} 
\vspace{10pt}
\caption{{\it ROSAT}~{\it PSPC} and {\it RXTE}~{\it PCA} observations
of the diffuse emission from G347.3-0.5 over the energy range 0.5 
through 30 keV.}
\label{fig1}
\end{figure}

\begin{table}
\caption{Best-Fit Parameters for {\it SRCUT} + {\it EQUIL} Model}
\begin{tabular}{lcccccc}
\hline
Section of & $n$$_H$ & & $\nu$$_{break}$\tablenotemark{a} & 
$K$\tablenotemark{a} & {\it kT}\tablenotemark{b} \\
G347.3-0.5 & (10$^{22}$ cm$^{-2}$) & $\alpha$\tablenotemark{a} & 
(10$^{17}$ Hz) & (Jy at 1 GHz)  & (keV)\\
\hline
NW Rim & 0.77$\pm$0.01 & 0.4996$\pm$0.0002 & 1.4578$\pm$0.005 & 
1.787$\pm$0.011 & 1.375$\pm$0.016 \\
SW Rim & 0.61$\pm$0.02 & 0.5143$\pm$0.0003 & 2.4235$\pm$0.013 & 
0.907$\pm$0.007 & '' \\
NE Region & 0.49$\pm$0.02 & 0.5341$\pm$0.0003 & 2.9269$\pm$0.021 & 
0.980$\pm$0.009 & '' \\
Diffuse & 0.51$\pm$0.01 & 0.5000$\pm$0.0002 & 2.2751$\pm$0.004 & 
1.989$\pm$0.005 & '' \\
\hline
\hline
\end{tabular}
\tablenotetext{a}{{\it SRCUT} fit parameter with single parameter
1$\sigma$ errors.} 
\tablenotetext{b}{{\it EQUIL} fit parameter with single parameter 
1$\sigma$ errors.}
\end{table}

\section{Results and Conclusions}

The results of this work may be summarized as follows:

1) Remarkably good fits (a reduced $\chi$$^2$ of 1.9 for 487 degrees
   of freedom for all of the data sets in Table 1)
   to the X-ray spectrum of G347.3-0.5 
   as observed by the $\it{ROSAT}$, $\it{ASCA}$ and $\it{RXTE}$
   satellites have been obtained by using a combination of
   non-thermal and thermal models (SRCUT and EQUIL) in 
   XSPEC (see Table 2).
   The X-ray spectrum of G347.3-0.5 at energies higher
   than 8 keV cannot be adequately fit by simply extending the
   power-law fits presented by Slane et al. 1999.

2) We have found evidence for modest thermal emission from this SNR.
   The energy of this thermal component is rather large ($\it{kT}$ 
   $\approx$ 1.4 keV) and appears to be associated with the diffuse 
   emission from the SNR rather than the X-ray luminous rims. At an
   assumed distance of 6 kpc, the corresponding ambient density is
   $n$$_H$ $\approx$ 0.02 cm$^{-3}$, consistent with the works of
   other authors (Slane et al. 1999, Ellison et al. 2001).

3) For the range of break frequencies determined for the different 
   portions of G347.3-0.5 (1.5 through 2.9 $\times$ 10$^{17}$ Hz), we 
   follow the example of Reynolds \& Keohane (1999) and calculate a 
   maximum electron energy of $\approx$ 70 TeV,
   assuming an ambient magnetic field strength of 10 $\mu$G. This 
   result is larger than the maximum electron energy calculated 
   with different models for G347.3-0.5 by Ellison et al. 2001.

\acknowledgments

The authors acknowledge support from NASA LTSA grant NAG5-9237.

\end{document}